\documentclass[english,aps,prl,superscriptaddress,twocolumn]{revtex4}
\usepackage{amsmath}
\usepackage{color}
\usepackage{graphicx}
\usepackage{amssymb}
\usepackage{yfonts}

\begin{document}

\title{Pressure-induced electronic mixing and enhancement of ferromagnetic ordering in EuX (X=Te, Se, S, O) magnetic semiconductors}

\author{Narcizo M. Souza-Neto}

\affiliation{Advanced Photon Source, Argonne National Laboratory, Argonne, IL
60439, U.S.A.}

\author{Daniel Haskel }

\affiliation{Advanced Photon Source, Argonne National Laboratory, Argonne, IL
60439, U.S.A.}

\author{Yuan-Chieh Tseng}

\affiliation{Department of Materials Science and Engineering, Northwestern University,
Evanston, IL 60201, U.S.A.}

\affiliation{Advanced Photon Source, Argonne National Laboratory, Argonne, IL
60439, U.S.A.}

\author{Gerard Lapertot}

\affiliation{Institut Nanosciences et Cryog\'enie, SPSMS,
CEA Grenoble, 38054 Grenoble, France}

\begin{abstract}
The pressure- and anion-dependent electronic structure of EuX (X=Te, Se, S, O) monochalcogenides is probed with element- and orbital-specific X-ray absorption spectroscopy in a diamond anvil cell. An isotropic lattice contraction enhances the ferromagnetic ordering temperature by inducing mixing of Eu 4{\it f} and 5{\it d} electronic orbitals. Anion substitution (Te $\rightarrow$ O) enhances competing exchange pathways through spin-polarized anion {\it p} states, counteracting the effect of the concomitant lattice contraction. The results have strong implications for efforts aimed at enhancing FM exchange interactions in thin films through interfacial strain or chemical substitutions.
\end{abstract}
\maketitle

Monochalcogenides EuX (X = Te, Se, S, O) materials \citep{Wachter-1979,MAUGER-PR1986} have attracted renewed interest  for displaying fully spin-polarized bands \citep{Steeneken-PRL2002,Schmehl-NatMat2007,Laan-PRL2008}, and colossal magnetoresistance (CMR) more pronounced than in manganite compounds \citep{Shapira-PRB73}, making them attractive for potential use in spintronic devices. Although the ferromagnetic (FM) ordering temperature, $\mathrm{T_{C}}$, is below 70 K in the bulk limiting its  applications, a lattice contraction induced by pressure \citep{Goncharenko-PRL1998,Kunes-JPSJ2005,Rupprecht-thesis}, chemical doping \citep{Ott-PRB2006,Wachter-1979,MAUGER-PR1986} or interfacial strain \citep{Lechner-PRL2005,Schmehl-NatMat2007,Ingle-PRB2008} significantly increases $\mathrm{T_{C}}$ toward room temperature. For example, EuS reaches $\mathrm{T_{C}} \approx$ 290 K at P = 88 GPa \citep{Rupprecht-thesis}. With successful reports of EuO integration with Si and GaN \citep{Schmehl-NatMat2007}, and that a biaxial lattice contraction increases $\mathrm{T_{C}}$ in epitaxially strained films \citep{Schmehl-NatMat2007,Lechner-PRL2005,Ingle-PRB2008}, the promise for EuX usage in spin-electronics is kept up. 
An understanding of the changes in electronic structure responsible for the strengthening of the indirect FM exchange interactions, while key to enable further developments, remains elusive with models alternatively centered on {\it f-d}, {\it s-f} or {\it p-f}  mixing being proposed \citep{Kasuya-IBM1970,NOLTING-PSSB1979,Liu-SSC1983}. 
While evidence for a significant role of Eu 4{\it f}-5{\it d} mixing was provided by optical spectroscopy \citep{Syassen-Physica86,Wachter-1979}, M\"ossbauer studies are masked by the dominant contribution of {\it s} shells to the isomer shift \citep{Klein-1976}, and neutron diffraction under pressure showed that a strong influence of the anion {\it p} shells must be considered \citep{Goncharenko-PRL1998}.
Additionally, the possible existence of mixed valent Eu at high pressures and its relation to the upturn in $\mathrm{T_{C}}$(P) observed in EuO after reaching 200 K at P $\approx10$ GPa, remains a matter of debate \cite{ABDELMEGUID-PRB1990,TISSEN-1987,Zimmer-PRB1984,Ingle-PRB2008}.
It is noteworthy that a direct probe of the element- and orbital-specific eletronic structure is still lacking, thwarting efforts to conclusively explain the effect of lattice contraction upon the mechanism of indirect exchange regulating $\mathrm{T_{C}}$ in these materials.

In this letter, we exploit the element- and orbital-selectivity of Eu {\it L}-edge x-ray absorption spectroscopy  in electric-dipole (2{\it p} $\rightarrow$ 5{\it d}) \citep{Schutz-XMCD} and electric-quadrupole (2{\it p} $\rightarrow$ 4{\it f}) \citep{Pettifer-Nat08, Lang-QuadXMCD} channels to probe the spin-polarized electronic structure of Eu 4{\it f} (valence) and 5{\it d} (conduction) states  as the lattice is contracted with chemical (Te $\rightarrow$ O) or physical pressure in a diamond anvil cell (DAC). 
Our results, supported by density functional theory (DFT), provide direct spectroscopic evidence that  Eu 4{\it f}-5{\it d} electronic mixing is dramatically enhanced under physical pressure and dictates the FM ordering temperature, $\mathrm{T_C}$. This mixing is much weaker when an equal lattice contraction is induced with anion substitution, a result of a concomitant increase in anion {\it p}   and Eu (4{\it f}, 5{\it d}) interactions that diminish the  strength of indirect FM exchange. 
 No mixed-valent Eu is observed within the pressure range of our measurements, and the upturn in $\mathrm{T_{C}}$ seen in EuO at high pressure \cite{ABDELMEGUID-PRB1990,TISSEN-1987} is attributed to the weakening of electron-electron correlations as the strongly hybridized Eu 5{\it d} states become delocalized at the insulator-metal transition \citep{Zimmer-PRB1984}. The results pinpoint the relevant changes in electronic structure regulating $\mathrm{T_C}$ in these materials, and should help guide efforts aimed at tailoring exchange interactions in monochalcogenide thin films through manipulation of interfacial strain \citep{Lechner-PRL2005} and chemical doping \citep{Ott-PRB2006}.

Polycrystalline samples of EuX  were prepared as described in Ref.~\citep{MAUGER-PR1986}. X-ray diffraction and SQUID magnetization measurements as a function of temperature show
good agreement with the known crystal structure and magnetic ordering temperatures \citep{Wachter-1979,MAUGER-PR1986}. X-ray absorption near edge structure (XANES) and X-ray magnetic circular dichroism (XMCD) experiments were carried out at beamline 4-ID-D of the Advanced Photon Source. The beamline is equipped with
phase-retarding optics to convert the linear polarization of
synchrotron radiation to circular \citep{Lang-RSI95}. The XMCD spectra are obtained from measurements of the helicity-dependent absorption coefficient, $\mu^{+,-}$, normalized by the absorption edge-jump, as ($\mu^{+}-\mu^{-}$) while the spin-averaged XANES is defined as ($\mu^{+}+\mu^{-}$)/2. XMCD measurements were carried out for two directions of the applied magnetic
field, along and opposite the incident photon wave vector, to check for systematic errors. Data were collected on powders in transmission geometry at ambient pressure (T = 5 K) and high-pressures (up to 15 GPa, T = 13 K). Pressure was calibrated {\it in-situ} by the Ruby luminescence method \citep{Syassen-HPR08}, and silicone oil was used as pressure-transmitting medium. X-ray transmission at the relatively low energy of the Eu L$_3$ edge (6.97 keV) was facilitated by using fully- and partially-perforated  
diamond anvils \citep{Dadashev-RSI01}. The diamond anvil cell (DAC) is in thermal contact with the cold finger of
a He flow cryostat, which is placed between the pole pieces of an  
electromagnet in a 0.5 T applied field \citep{Haskel-RSI07}. DFT {\it ab initio} calculations were performed using the WIEN2k implementation of the full-potential
linearized augmented-plane-wave (FLAPW) method \citep{Blaha-Wien2k}. The LDA+U calculations use similar parameters to those previously reported for this system \citep{Kunes-JPSJ2005,Ingle-PRB2008}, including U=7eV.  Only ferromagnetic structures were considered and spin-orbit coupling was found to be negligible.

\begin{figure}
\begin{centering}
\includegraphics[scale=0.165]{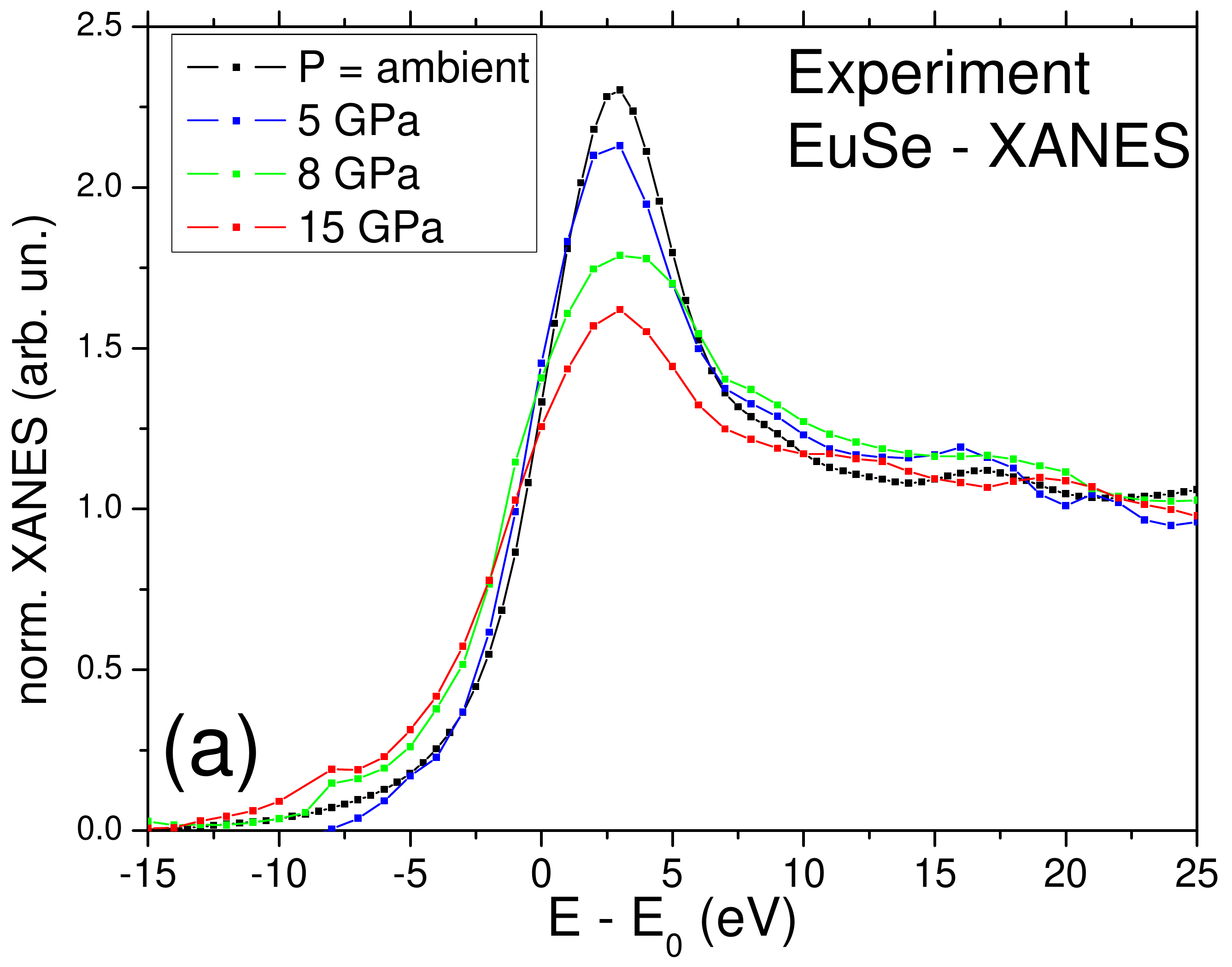}~
\includegraphics[scale=0.165]{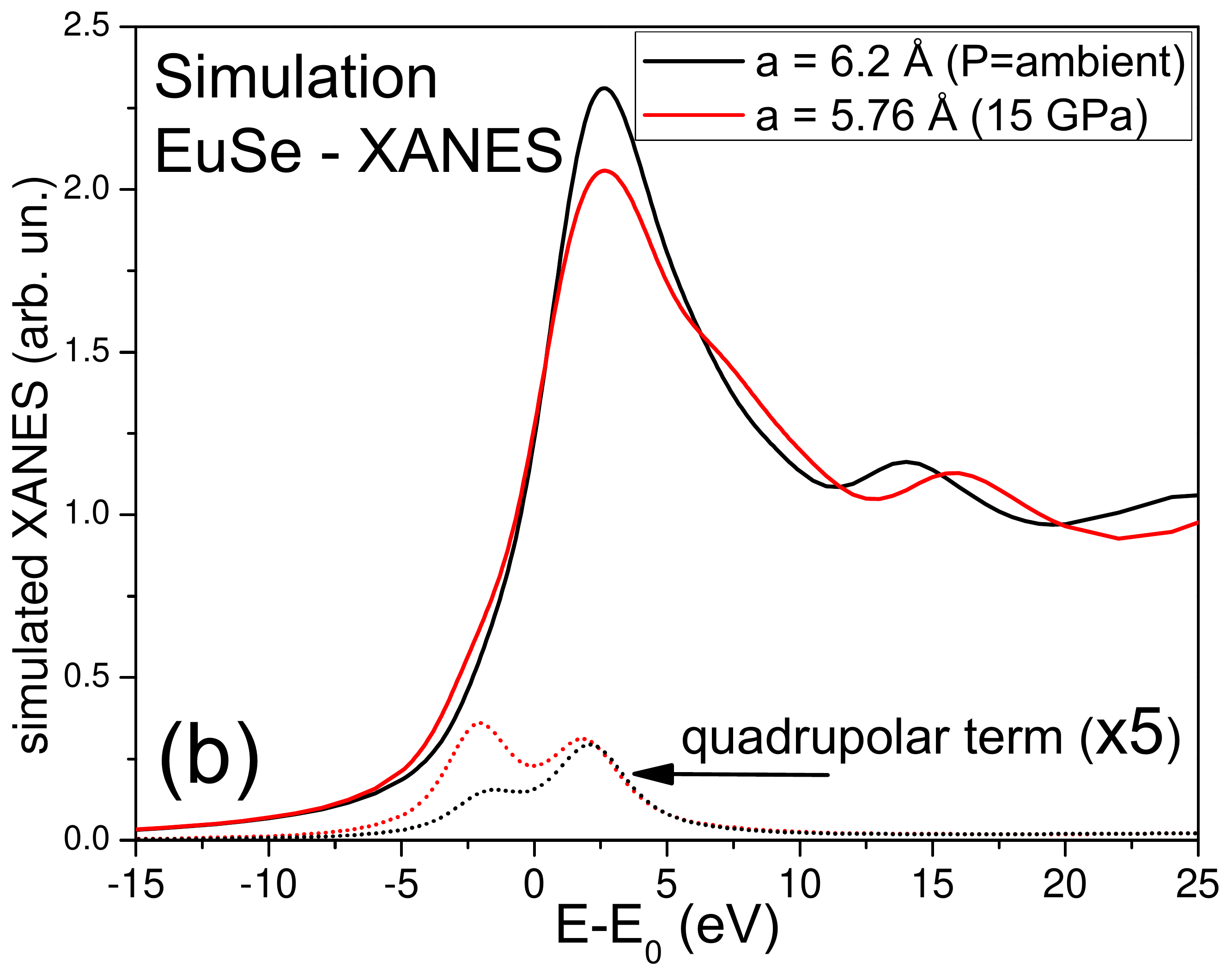}
\par\end{centering}

\caption{\label{fig:xanes-EuSe} (Color online) Experimental (a) and simulated (b)
XANES spectra for EuSe as a function of applied pressure. The two peaks in the quadrupolar contribution
come from majority and minority 4{\it f} states, separated by
$\approx4.5\,\mathrm{eV}$. The energy is referenced to the $\mathrm{Eu^{2+}}$
$\mathrm{L_{3}}$ absorption edge at 6.970 keV.}

\end{figure}

\begin{figure}
\begin{centering}
\includegraphics[scale=0.155]{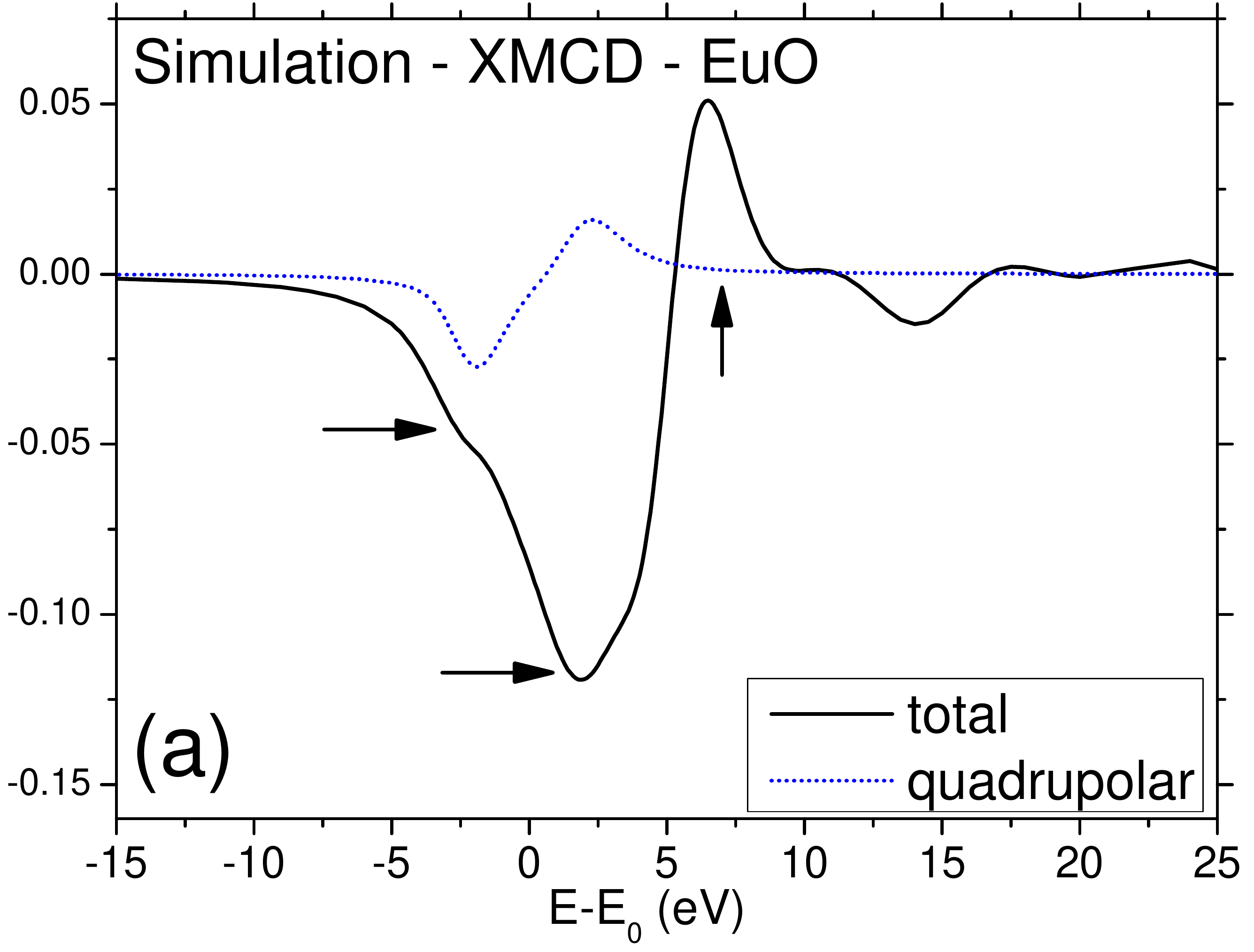}\includegraphics[scale=0.155]{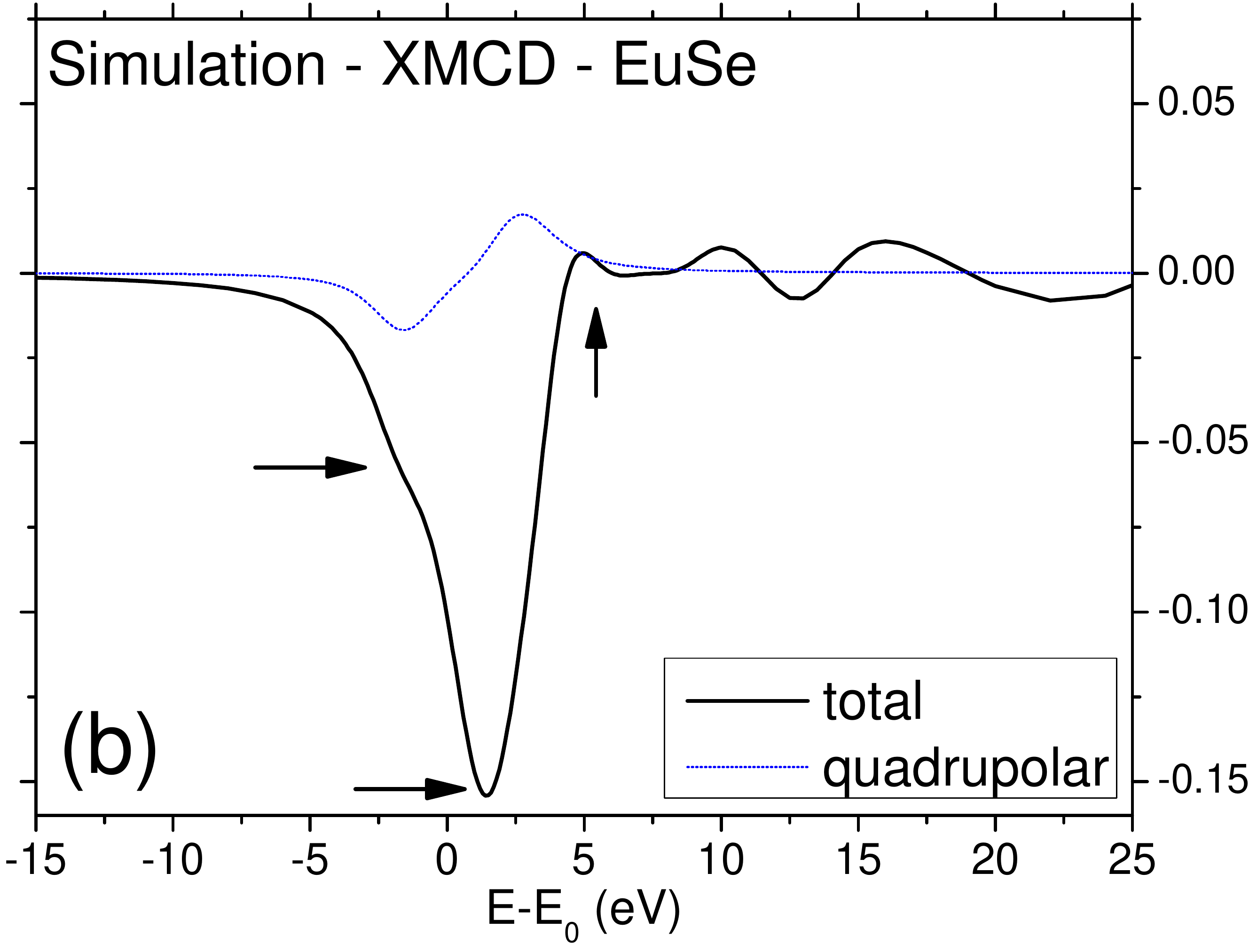}
\par\end{centering}

~

\begin{centering}
\includegraphics[scale=0.165]{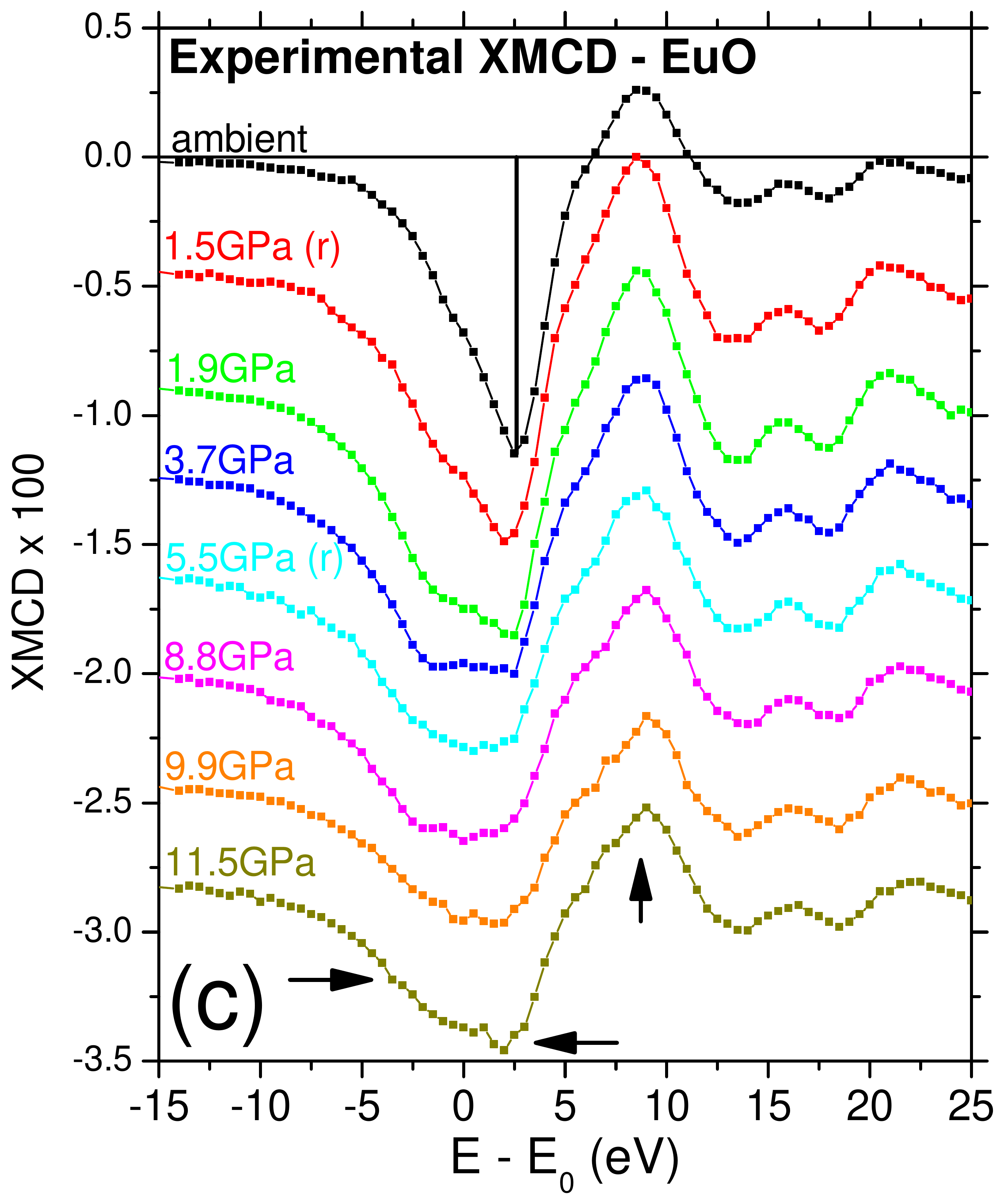}\includegraphics[scale=0.165]{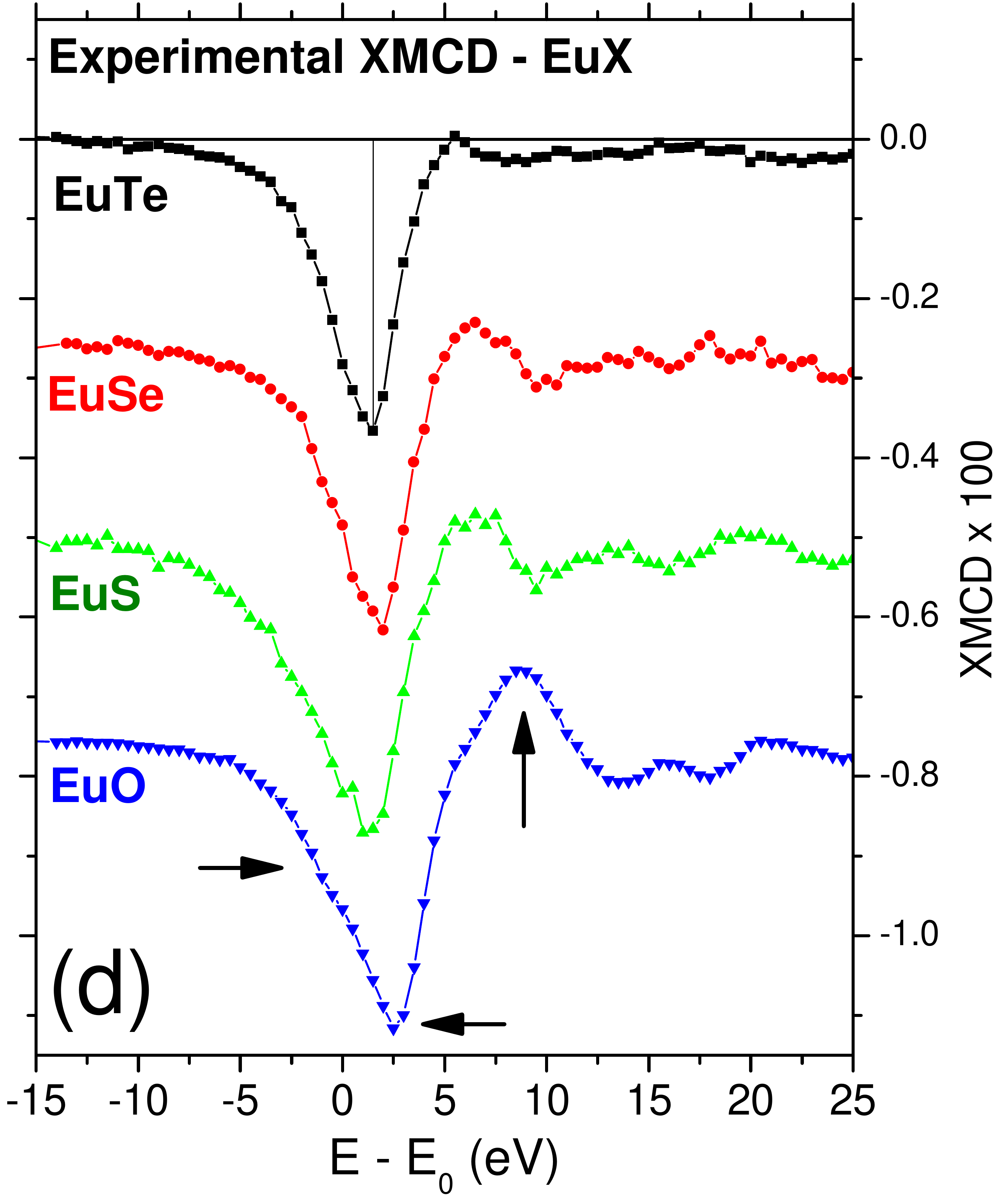}
\par\end{centering}

\caption{\label{fig:xmcd-EuX-EuO} (Color online). FDMNES simulations
of XMCD spectra for (a) EuO and (b) EuSe compounds. (c) Pressure-dependent XMCD data for EuO at T=13 K. Data are shifted
vertically for clarity.  Data taken during pressure release (r) shows lack of significant hysteresis. Arrows indicate spectral features with  
Eu 4{\it f}, Eu 5{\it d} and anion-p orbital character. (d) XMCD spectra for EuX at T = 5 K. Data are normalized to the peak amplitude of EuSe and shifted vertically for clarity. } 

\end{figure}

\begin{figure}
\begin{centering}
\includegraphics[viewport=40bp 70bp 580bp 425bp,clip,scale=0.46]{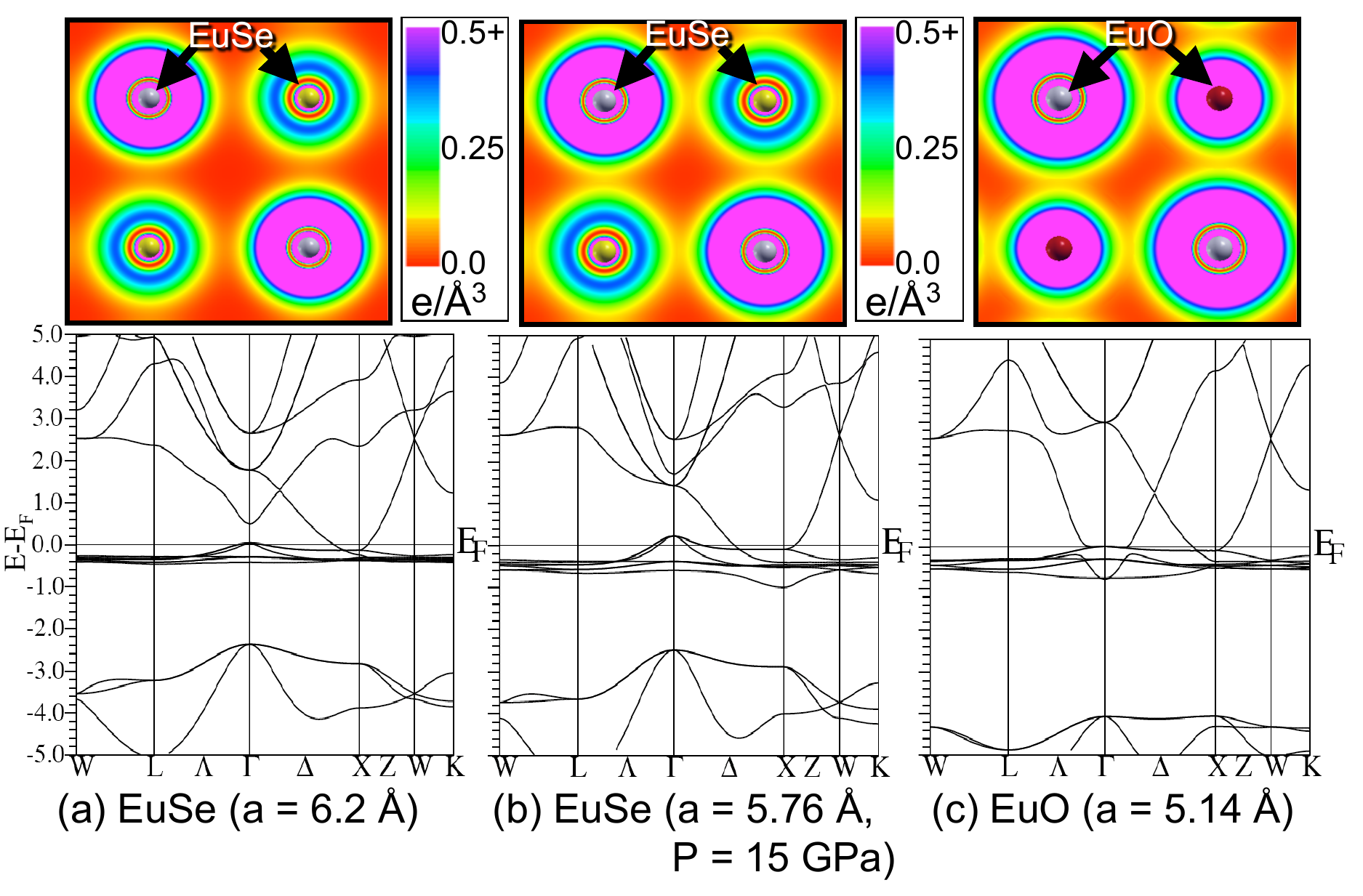}
\par\end{centering}

\begin{centering}
\includegraphics[scale=0.158]{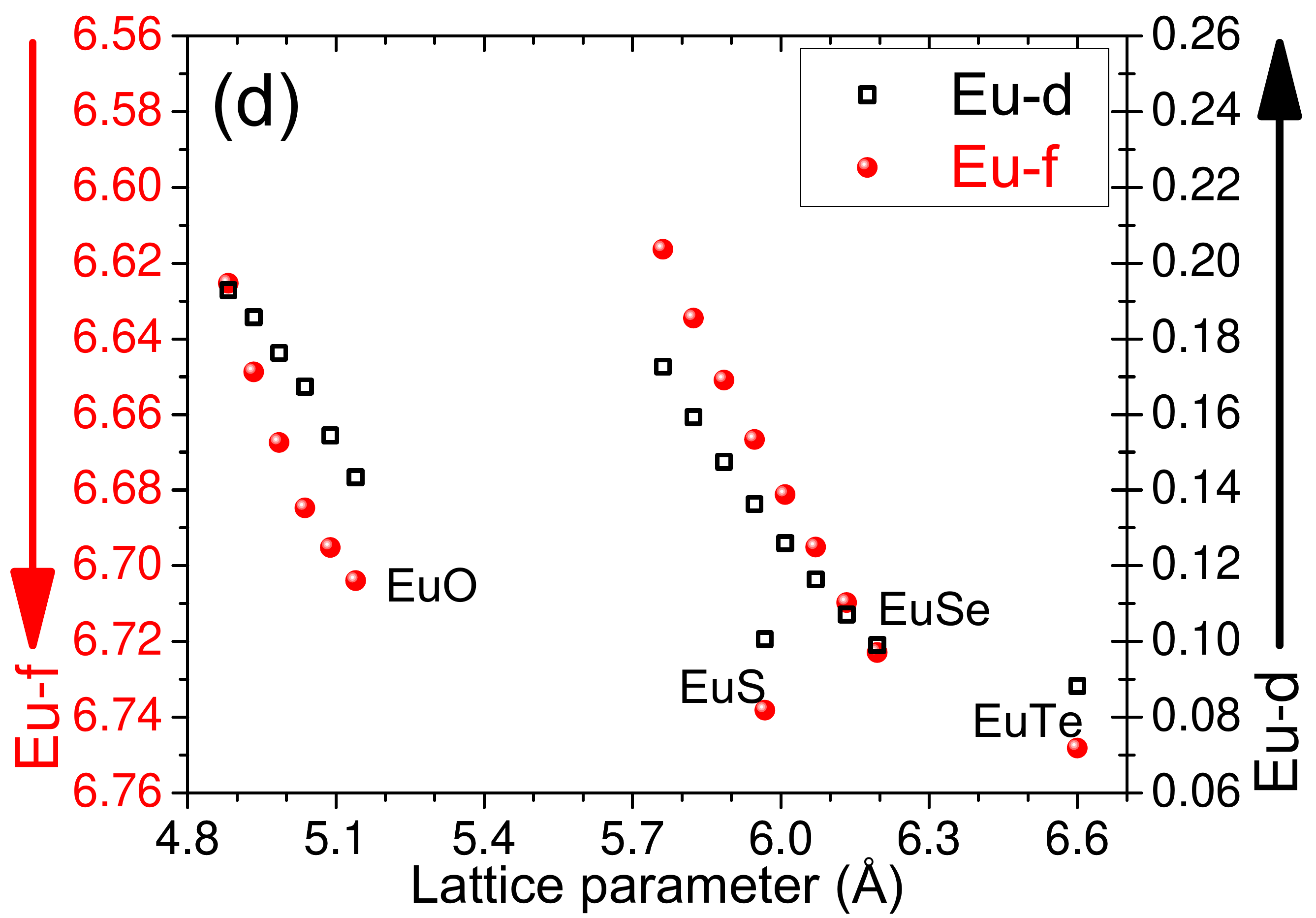}
~\includegraphics[scale=0.153]{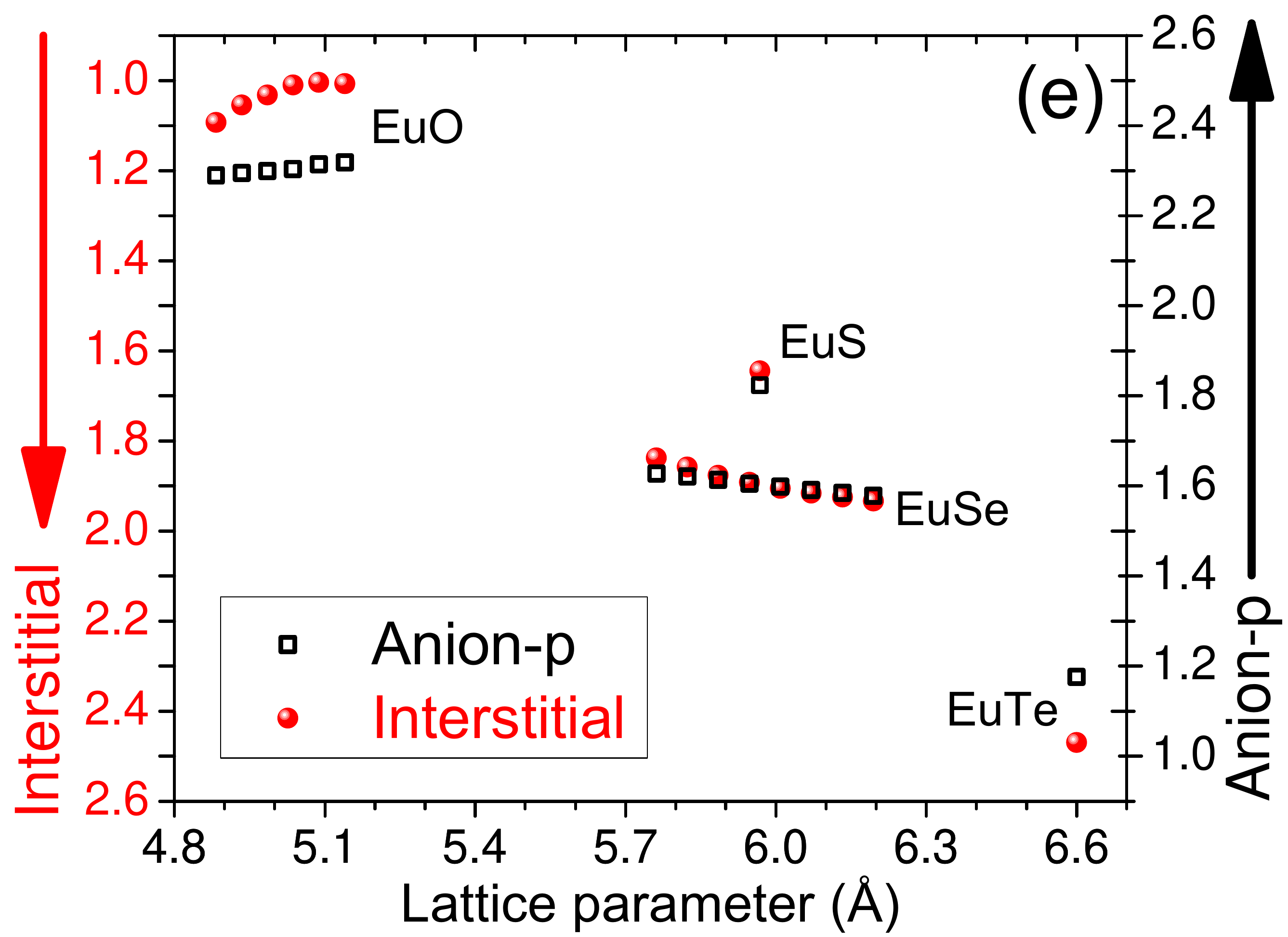} 
\par\end{centering}

\caption{\label{fig:dft-results} (Color online) Majority (spin-up) bands and charge density maps obtained from DFT calculations in the ferromagnetic state of EuSe at ambient pressure (a) and 15 GPa (b) and EuO at ambient pressure (c).
Anion {\it p} character dominates the valence band while Eu 5{\it d}
dominates the conduction band. Occupied Eu 4 {\it f} states  
are localized just below (or crossing) the Fermi level inside the semiconducting gap (minority 4{\it f}
orbitals are at $\approx4.5$eV and not shown). Majority electron occupancies for Eu 4{\it f} and Eu 5{\it d} states are shown in (d) while changes in interstitial and anion-{\it p} occupation are shown in (e).}

\end{figure}

Figure~\ref{fig:xanes-EuSe} shows representative pressure-dependent XANES data for EuSe and Figure~\ref{fig:xmcd-EuX-EuO} shows pressure- (EuO) and anion-dependent XMCD data, all taken at the Eu L$_3$ absorption edge
(2{\it p}$_{3/2}$ initial state) of these compounds. Small amounts of Eu$_2$O$_3$ impurities ($<$ 3\%) were detected in EuO and EuTe samples. However, this non-magnetic impurity is invisible to XMCD, as verified
by null XMCD measurements on Eu$_2$O$_3$ powders (Eu$^{3+}$ is a J=0
ion). XANES and XMCD simulations carried out with the FDMNES code \citep{Joly-PRB01} are also shown. It is expected \citep{Rogalev-EuX} that a weak quadrupolar (2{\it p}$\rightarrow$ 4{\it f}) contribution dominates the photo-excitation process at the onset of x-ray absorption in these compounds, while dipolar contributions (2{\it p} $\rightarrow$ 5{\it d}) dominate at higher excitation energies. This assertion is confirmed by the FDMNES simulations, which facilitate assignment of XANES and XMCD spectral features to quadrupolar and dipolar channels. The lattice contraction induced by pressure results in a dramatic {\em decrease} in the amplitude of the main peak in XANES and XMCD spectra (also referred to as the white line), which is dominated by dipolar contributions and probes the density of empty Eu 5{\it d} states in the vicinity of the Fermi level. Oppposite to that, a concomitant {\em increase} in XANES and XMCD spectral weight is observed at lower energies where quadrupolar contributions dominate. This increase, which manifests an asymmetric broadening of the XMCD signal as the lattice is contracted
with applied pressure or with anion substitution [Figs.~\ref{fig:xmcd-EuX-EuO}(c-d)], is related to an  increase in empty density of 4{\it f} states. The spectral weight transfer from high to low energies observed in both XANES and XMCD data is
evidence that charge transfer from Eu 4{\it f} to Eu 5{\it d} states is taking place. Furthermore, the XMCD data unequivocally demonstrate that the newly formed empty 4{\it f} states are spin-polarized. DFT band structure calculations carried out at ambient- and high-pressure (contracted lattices), summarized in Figure~\ref{fig:dft-results}, support this interpretation. The calculations show that these empty 4{\it f} states are derived from spin-up majority states that reside below, but in close proximity to, the Fermi level at ambient pressure and large lattice volumes [Fig.~\ref{fig:dft-results}(a)], but become chemically active through hybridization with otherwise empty 5{\it d} states in the conduction band as the lattice is contracted [Fig.~\ref{fig:dft-results}(b)]. Electron occupations of Eu 4{\it f} and 5{\it d} majority orbitals are shown in Fig.~\ref{fig:dft-results}(d), with the majority charge in Eu 5{\it d} states increasing by about 0.1 e$^-$ while the charge on Eu 4{\it f} states decreases by the same amount over the entire change in lattice volume probed in the current experiments.

\begin{figure}
\begin{centering}
\includegraphics[scale=0.32]{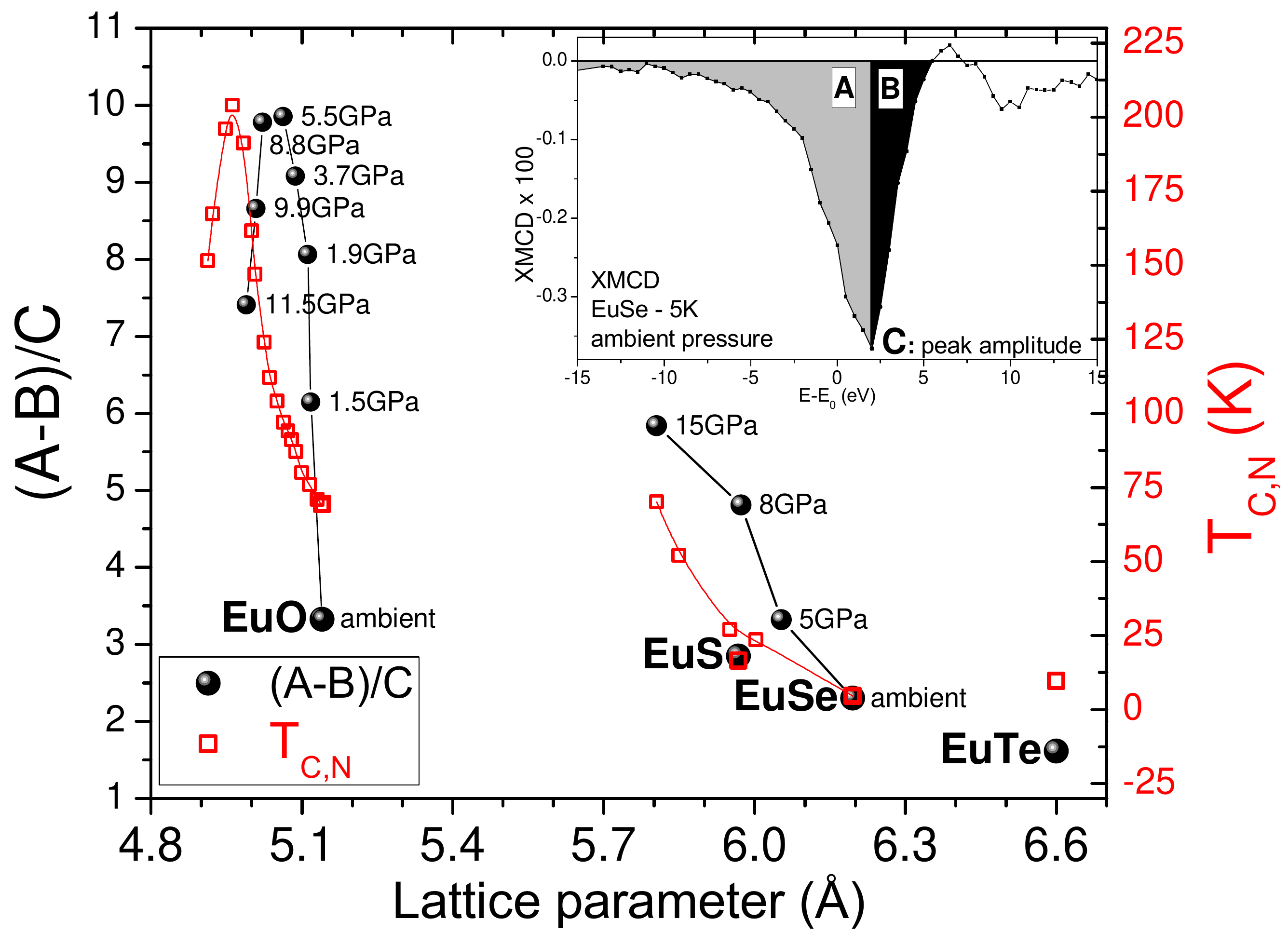}
\par\end{centering}

\caption{\label{fig:asymmetry-ratio-Tc} (Color online) Asymmetry in
XMCD signal defined as (A-B)/C.  A and B areas are defined in the inset and
C is XMCD peak amplitude. (A-B)/C is plotted in the main panel for EuX (X = 0, S, Se,
Te) at ambient pressure and EuO and EuSe at applied pressure. The magnetic
ordering temperatures $\mathrm{T_{C,N}}$, taken from Refs.~\citep{Goncharenko-PRL1998,TISSEN-1987},
are shown for comparison. The pressure-volume relationship was assumed to be the same 
as in \citep{Goncharenko-PRL1998} and references therein.}

\end{figure}

To quantify the asymmetric broadening of the dichroic signal arising from the enhancement (reduction) of quadrupolar (dipolar) contributions to the absorption spectra upon lattice contraction, we computed the mixing fraction (A-B)/C, defined in the inset of Fig.~\ref{fig:asymmetry-ratio-Tc}. This fraction grows for contracted lattices as the 4{\it f} states cross the Fermi level and become chemically active. The striking correlation between the mixing fraction and the magnetic ordering temperature $\mathrm{T_{C,N}}$ of these compounds \citep{Goncharenko-PRL1998} (Fig.~\ref{fig:asymmetry-ratio-Tc}), together with the calculated {\it f}-to-{\it d} charge transfer [Fig.~\ref{fig:dft-results}(d)], clearly demonstrates that Eu {\it f-d} mixing regulates the strength of indirect FM exchange in these compounds. Although {\it f-p} mixing also occurs \citep{Kunes-JPSJ2005}, we did not find a correlation between this mixing and the pressure-induced increase in $\mathrm{T_C}$. 
While both chemical- and physical-pressures result in enhanced {\it f-d} mixing and correspondingly larger FM ordering temperatures, it is clear from Fig.~\ref{fig:asymmetry-ratio-Tc} that anion substitution is much less effective than applied pressure in enhancing $\mathrm{T_{C,N}}$. For example, the FM ordering temperature of EuSe at 15 GPa surpasses that of EuO at ambient pressure, despite the much smaller lattice volume of the latter. In what follows we discuss possible reasons for this observation.

Contracting the lattice with chemical pressure (anion substitution Te $\rightarrow$ O) not only induces Eu {\it f-d} mixing as with physical pressure (albeit to a much lesser extent), but also causes other pronounced changes in the spin-polarized electronic structure. This is clearly seen in Fig.~\ref{fig:xmcd-EuX-EuO}(d) where a strongly X- and energy-dependent feature is observed in the XMCD spectra at $\approx 5-8$ eV above the edge.  DFT calculations of projected density of states (not shown but equivalent to those presented in Ref. ~\citep{Ingle-PRB2008}), show that this feature arises from empty {\it p} anion states, which acquire spin-polarization through hybridization with Eu {\it f,d} states. This anion-related feature systematically increases for Te $\rightarrow$ O but is largely unaffected by applied pressure (i.e. lattice contraction). This demonstrates that {\it f-d-p} hybridization is only weakly dependent on lattice contraction but strongly affected by anion type, e.g., through changes in crystal field splitting of Eu 5{\it d} states \citep{Goncharenko-PRL1998}. DFT calculations show that anion substitution (Te $\rightarrow$ O) leads to a very significant transfer of charge from the interstitial region to anion {\it p}-states, of about 1.2 (majority) e$^{-}$ over the entire volume change probed in our experiments [Figure~\ref{fig:dft-results}(e)]. This results in increased ionicity \citep{Gerth-PL68} of the rocksalt structure, as seen in the DFT charge density maps [Fig.~\ref{fig:dft-results}(a-c)]. While the lattice contraction enhances FM $\mathrm{T_C}$ by promoting Eu {\it f-d} mixing, a competing exchange pathway mediated by spin-polarized anion {\it p} states \citep{Rogalev-EuX} becomes more predominant for Te $\rightarrow$ O and counteracts the effect of the lattice contraction, limiting the $\mathrm{T_{C}}$ of EuO at ambient pressure to below 70 K.

The observed reduction in $\mathrm{T_C}$ for EuO at high pressures is still a matter of debate \cite{ABDELMEGUID-PRB1990,TISSEN-1987,Zimmer-PRB1984,Ingle-PRB2008}. Although pressure-induced, non-magnetic $\mathrm{Eu^{3+}}$ states \citep{Zimmer-PRB1984,Ingle-PRB2008} could result in a reduction of $\mathrm{T_C}$, we found no evidence for mixed valence behavior in the pressure range discussed here (P $<$ 15 GPa) (within the $\approx3$ at.\% accuracy of our measurements), in agreement with Ref. \cite{ABDELMEGUID-PRB1990}. A correlated decrease is observed in the (A-B)/C mixing fraction for pressures near where the upturn of Tc occurs in EuO, due to a small  increase of the  XMCD peak amplitude at high pressures. This may be interpreted as a signature of electron delocalization as a result of reduced electron-electron correlations with increasing {\it f-d} hybridization.  Interestingly, the occurrence of an insulator-metal (I-M) transition associated with the closing of the 4{\it f}-5{\it d} gap has been reported for EuO for pressures where the upturn in T$_{\rm C}$ is observed \citep{Zimmer-PRB1984,Ingle-PRB2008}. It is likely that the critical limit for $\mathrm{T_C}$ in EuO is reached when a strong {\it f-d} mixing forces the hybridized {\it d} states to become delocalized. We note that mixing of Eu {\it f-d} states already occurs at low pressures, so the reported I-M transition \cite{Zimmer-PRB1984} is not coincident with the onset of such mixing, contrary to the arguments in Ref.  \citep{Ingle-PRB2008}.  One would expect that an upper limit in $\mathrm{T_C}$ exists for the other EuX compounds as well, but at much higher pressures yet to be achieved \citep{Rupprecht-thesis}.

A theoretical understanding of the relevant interactions mediating indirect exchange in these materials is far from complete. Kasuya \citep{Kasuya-IBM1970} proposed that the ferromagnetism in this system is mediated by virtual excitations of 4{\it f} electrons into  5{\it d} states without any significant role for the anion, while Liu \citep{Liu-SSC1983} highlights the relevance of  spin-polarized anion {\it p} states and their interaction with localized {\it f} electrons. Our results demonstrate that a lattice contraction alone increases FM T$_{\rm C}$ by virtue of {\it f-d} mixing without significant involvement of anion {\it p} states. However, the role of anion {\it p} states is critical in explaining the effects of chemical pressure exerted by anion substitution. 

Future experimental work aimed at controlling and enhancing FM T$_{\rm C}$ in epitaxial films with interfacial strain or chemical substitutions should benefit from the results presented here. Although biaxial strain is less effective than hydrostatic pressure in increasing T$_{\rm C}$ due to concomitant in-plane and out-of-plane lattice modifications in thin films \citep{Lechner-PRL2005,Ingle-PRB2008}, magnetic superlattices may overcome this difficulty \citep{Lechner-PRL2005}. Furthermore, promoting electron localization by symmetry-breaking  at the interfaces of strained films \citep{Freeland-NM} may allow achieving  FM ordering temperatures beyond those corresponding to the electron-delocalization limit in bulk EuO.

\begin{acknowledgments}
The authors are grateful to Michael Norman, Michel van Veenendaal, Mark Antonio and Yves
Joly for discussions and comments. Work at Argonne is supported by the U.S. Department of Energy, Office of Science, Office of Basic Energy Sciences, under Contract No. DE-AC-02-06CH11357. 
\end{acknowledgments}

\bibliographystyle{apsrev}

\end{document}